\documentclass[aps,prl,twocolumn,superscriptaddress,showpacs,amsmath]{revtex4}

\usepackage{graphicx}

\begin{document}

\title{Field-induced chiral phase in isotropic frustrated spin chains} 

\author{Alexei Kolezhuk}
\thanks{On leave from the Institute of Magnetism, Academy of Sciences and
  Ministry of Education, 03142
  Kiev, Ukraine}
\affiliation{Institut f\"ur Theoretische Physik, Universit\"at
Hannover, Appelstra{\ss}e 2, 30167 Hannover, Germany}

\author{Temo Vekua}
\thanks{On leave from the Andronikashvili Institute of Physics, 380077 Tbilisi, Georgia}
\affiliation{Universit\'{e} Louis Pasteur,
Laboratoire de Physique Th\'{e}orique,
3 Rue de l'Universit{\'e}, 67084 Strasbourg C\'edex, France}

\begin{abstract}
It is shown that an external magnetic field applied to a spin-$S$ isotropic
zigzag chain induces a phase with spontaneously broken parity, characterized by
long range ordering of vector chirality. This is in contrast to the
two-component Luttinger liquid scenario proposed in the literature. Relevance to
real materials is discussed.
\end{abstract}

\maketitle

\emph{Introduction.--} In recent years, phases with broken vector chirality in
quantum spin chains have attracted a considerable interest
\cite{Nersesyan+98,Kaburagi+99,K00,Lecheminant+01,Hikihara+01}.  They are
characterized by a nonzero expectation value of the vector product of two
adjacent spins $\vec{\kappa}_{n} = \langle \vec{S}_{n}\times\vec{S}_{n+1}
\rangle$, so that in a chirally ordered phase spins tend to ``rotate'' in a certain
preferred plane predominantly clockwise or counterclockwise. This type of order
breaks only a discrete symmetry between left and right and thus is allowed in
one dimension (1D), in contrast to the long-range helical spin order
\cite{Villain78,Chubukov91}.  Recently, chirally ordered phases were numerically found in
frustrated chains with easy-plane anisotropy \cite{Kaburagi+99,Hikihara+01}.
The chiral ordering transition has possibly been observed experimentally in the
quasi-1D anisotropic organic magnet $\rm Gd(hfac)_{3}NITiPr$
\cite{Affronte+99}.

In this Letter we show that a chiral phase emerges in
\emph{isotropic} frustrated spin
chains as well, if they are subject to a strong external magnetic field.
We focus on the model of a zigzag chain defined by the Hamiltonian:
\begin{equation}
\label{Theham}
{\mathcal H}=J_{1}\sum_{j} \vec{S}_{j}\cdot \vec{S}_{j+1} 
+J_{2}\sum_{j}\vec{S}_{j}\cdot \vec{S}_{j+2} -h \sum_{j}S_{j}^{z}
\end{equation}
where $\vec{S}_j$ are spin-$S$ operators at the $j$-th site, and $J_{1,2}>0$.
It is easy to analyse the \emph{classical} counterpart of the above model,
where spins are represented by vectors, $(S_n^{\pm},S_{n}^{z})
\mapsto(S\sin \phi_n e^{\pm i\theta_n},S\cos\phi_n)$. The applied field
selects a preferred plane, reducing the symmetry to $\rm U(1)$. Depending on the
frustration strength $\alpha=J_{2}/J_{1}$,
the in-plane ground state configuration is given by $\theta_{n}=(\pi\pm \lambda)n$, 
with $\lambda=0$ for $\alpha<1/4$ and $\lambda=\arccos(1/4\alpha)$ for
$\alpha>1/4$, respectively. The spins are canted towards the field,
$\cos\phi_{n}=h/h_{s}$, where
$h_{s}=4S\{J_{1}\cos^{2}(\lambda/2)+J_{2}\sin^{2}\lambda\}$ is the
saturation field. The classical ground state is a canted
antiferromagnet for $\alpha$ below the  Lifshits point $\frac{1}{4}$,
while for $\alpha>\frac{1}{4}$ one has two degenerate helical ground states, 
as reflected by
the $\pm$ signs above, which correspond to the left and right
chirality $\kappa=\pm S^{2}\sin\lambda$.  Thus, for
$\alpha>\frac{1}{4}$ in presence of a field the initial $\rm
SU(2)$ symmetry is reduced to $\rm U(1)\times Z_{2}$. 

In the \emph{quantum} case the $\rm U(1)$ symmetry cannot be
broken, but it is allowed to break the discrete $Z_{2}$ chiral symmetry.  Such
a scenario is realized in \emph{anisotropic} chains, where two different chiral
phases were found
\cite{Nersesyan+98,Kaburagi+99,Hikihara+01,Lecheminant+01,K00}. 
A natural question arises: Can an external magnetic field
act similarly to the $xy$ anisotropy \cite{khymyn}, favoring the chiral order
in \emph{isotropic} spin chains?  Recent numerical studies
\cite{Okunishi+99,OkunishiTonegawa03} propose
the scenario of a \emph{two-component Luttinger liquid} without any
breaking of the $Z_{2}$ symmetry, casting doubts on the above
idea.  The aim of the present Letter is to show that the correct
high-field physics of isotropic frustrated chains is indeed determined by the
spontaneous breaking of the chiral symmetry.

\emph{$S=\frac{1}{2}$ chain.--} We start with the extreme quantum spin-$\frac{1}{2}$ case which
admits a field-theoretical description based on the bosonization
approach. Consider the limit of strong frustration $\alpha\gg 1$ and strong
magnetic fields $h\sim J_2$.  The system may be viewed as two chains weakly
coupled by the zigzag interaction $J_{1}$.  A single spin-$\frac{1}{2}$
chain in uniform magnetic field is known to be
critical, its low-energy physics being effectively described by the standard
Gaussian theory \cite{LutherPeschel} known also as the Tomonaga-Luttinger
liquid:
\begin{equation}
\label{SpinChainBosHam}
{\cal H} =  \frac{v}{2}\int dx \, \Big\{\frac{1}{K}(\partial_x \phi)^{2} 
+ K (\partial_x \theta)^{2}\Big\}.
\end{equation}
Here $\phi$ is a compactified scalar bosonic field and $\theta$ is its
dual, $\partial_t \phi =v \partial_x \theta $, with the commutation
relations $[\phi(x),\theta(y)] = i\Theta (y-x)$, where $\Theta(x)$ is the
Heaviside function and the regularization $[\phi(x),\theta(x)] =i/2$ is assumed.
Integrability of the $S=\frac{1}{2}$ chain model makes possible to relate
explicitly the coupling constants of the theory, the spin wave velocity $v$ and
the Luttinger liquid (LL) parameter $K$, to the microscopic parameters $J_{2}$,
$h$.  The exact functional dependences $v(h)$
and $K(h)$ are known (see
\cite{AffleckOshikawa} and references therein) from the numerical solution   of
the Bethe ansatz integral equations \cite{Bogoliubov}.
 Particularly, $K$ increases
with the magnetic field from $K(h=0)=\frac{1}{2}$ to $K=1$ for $h$
approaching the saturation value $2J_2$.  

In the infrared limit the
following representation of the lattice spin operators holds \cite{LutherPeschel}:
\begin{eqnarray}
\label{Luther}
S^z_n&=&\frac{1}{\sqrt{\pi}}\partial_x \phi
+\frac{a}{\pi} \sin \big\{2k_Fx+ \sqrt{4\pi } \phi\big\}  +m\\
S_n^-& =& (-1)^n e^{-i\theta\sqrt{\pi}}
\big\{ c+b\sin{\big(2k_F x+ \sqrt{4\pi }\phi\big)}\big\},\nonumber
\end{eqnarray}
Here $m(h)$ is the ground state magnetization per
spin which determines the Fermi wave vector $k_F=(\frac{1}{2}-m)\pi$ and is
known exactly from the Bethe ansatz results \cite{Bogoliubov}.  Nonuniversal
constants $a$, $b$, and $c$ for general $h$ have been extracted numerically 
from the density matrix renormalization group (DMRG) calculations \cite{Furusaki}.

We treat the $J_{1}$  interchain
coupling term  perturbatively, representing two decoupled
chains in terms of
Gaussian models of the form (\ref{SpinChainBosHam}). 
 It is convenient to pass to the symmetric and
antisymmetric combinations of the fields describing the individual chains,
$\phi_{\pm}=(\phi_1\pm \phi_2)/\sqrt{2K}$ and  
$\theta_{\pm}= (\theta_1 \pm \theta_2)\sqrt{K/2}$. 
The effective Hamiltonian describing low-energy properties of the 
model (\ref{Theham}) takes the following form:
\begin{eqnarray}
\label{symantisym}
{\mathcal H}_{\rm eff}  &=& {\mathcal H}_{0}^{+} +{\mathcal H}_{0}^{-} +
{\mathcal H}_{\rm int},\nonumber\\
{\mathcal H}_{0}^{\pm} & =&\frac{v}{2} [(\partial_x \theta_{\pm})^{2} + (\partial_x
  \phi_{\pm})^2], \nonumber\\
{\mathcal H}_{\rm int} &=& g_1\cos{k_F}\cos\big(k_F+\sqrt{8\pi K_-}\phi_{-}\big) 
\nonumber\\
&-& g_2\partial_x \theta_+\sin\big(\sqrt{2\pi/K_{-}}\theta_-\big).
\end{eqnarray}
Only the relevant terms are shown here \cite{comment1}, 
including the ``twist operator'' with
nonzero conformal spin \cite{Nersesyan+98}. 
The Fermi velocity $v \propto J_{2}$, while
the couplings
$g_{0,1,2}\propto J_1\ll v$. The renormalized LL parameter is given by
\begin{equation}
K_-=K(h)\Big\{ 1 + J_1 K(h)/\big(\pi v(h)\big) \Big\}.
\end{equation}
Note that in the first order of $J_1/J_2$ the correction to $K_{-}$ for the
zigzag type of interchain coupling is twice larger compared to that for the
ladder type of coupling. 

The inter-sector part of (\ref{symantisym}) contains a term which can be
identified as an infrared limit of the
product of in-chain and interchain chiralities:  one can show that
\begin{equation}
\label{chirprod}
\partial_x \theta_+\sin\sqrt {\frac{2\pi}{ K_-}}\theta_- \propto
(\kappa^{z}_{2i-1,2i+1}+ \kappa^{z}_{2i,2i+2})\,\kappa^{z}_{2i,2i+1},
\end{equation}
where $\kappa_{i,j}^{z}\equiv (\vec{S}_{i}\times \vec{S}_{j})^{z}$.
All the other terms, omitted in (\ref{symantisym}), are made either irrelevant
or incomensurate by the external magnetic field. The Hamiltonian
(\ref{symantisym}) gives the minimal effective field theory 
 describing the low-energy dynamics of a strongly 
frustrated ($\alpha\gg1 $) spin-$\frac{1}{2}$ zigzag chain for a nonzero
magnetization $m$.  For small $m$ the LL parameter $K_{-}\simeq \frac{1}{2}$, and
the inter-sector $g_{2}$ term has a higher scaling dimension than the strongly
relevant  $g_{1}$ term in the antisymmetric
sector. In this case the system is in a phase with relevant coupling in
antisymmetric sector, as discussed for the first time 
in Ref.\ \cite{Cabra} (later dubbed EO phase
\cite{OkunishiTonegawa03}). In contrast to that, at $h=0$ all terms generated 
by the zigzag coupling are only marginal.

When $h$ increases, the chirality product operator
(\ref{chirprod}) can become more relevant than the $g_{1}$ term controling
the field $\phi_{-}$; the latter term  becomes less
relevant with the increase of $h$ as well as with the increase of the
zigzag \emph{antiferromagnetic} coupling $J_{1}$.  To study this situation, one
can apply a mean field decoupling procedure to the inter-sector term in the
spirit of Ref.\ \cite{Nersesyan+98}.  At the mean field level, the interaction
${\mathcal H}_{\rm int}$ takes the form
\begin{eqnarray}
\label{MF}
&&{\mathcal H}_{MF}= g_1\cos{k_F}\cos({k_F+\sqrt{8\pi K_-}\phi_-})\\
&& - g_2\partial_x \theta_+\langle\sin \sqrt{\tfrac{2\pi}{K_-}}\theta_-\rangle
- g_2\langle\partial_x \theta_{+}\rangle \sin\sqrt{\tfrac{2\pi}{K_-}}\theta_{-}.
\nonumber
\end{eqnarray}
Remarkably, the
mean field Hamiltonian reveals a competition between the basic and dual field
terms of the form $\sin(\gamma\phi_{-})$ and $\sin(\delta \theta_{-})$ with
$\gamma\delta=4\pi$, exactly the value where the Ising quantum phase transition
takes place \cite{Totsuka,Arlego}.
To find the critical magnetic field $h_{cr}$ which corresponds to this
transition, we equate the
RG masses produced by the operators $\sin(\gamma\phi_{-})$ and $\sin(\delta
\theta_{-})$,  substituting the  averages in
(\ref{MF}) by their mean-field values
found from self-consistency conditions. Doing so,  one obtains
an estimate for the renormalized LL parameter $K_{-}$ at the transition,
which happens to be related to
the glorious ``golden mean'' $q$:
\[
2K_{-}(h_{cr})=q \equiv(\sqrt{5}+1)/2.
\]
This leads to the following  equation for $h_{cr}$:
\begin{equation}
\label{critical}
K(h_{cr})=   \frac{q}{2} \Big\{ 1 - \frac{J_1 K(h_{cr})}{\pi v(h_{cr})} \Big\}
\end{equation}
The fact that $K(h)$ is a monotonically increasing function
 \cite{AffleckOshikawa,Honecker} 
 implies that
the critical field decreases with increasing the \emph{antiferromagnetic} zigzag
coupling $J_1$:
\begin{equation}
\label{importantresult}
(\partial h_{cr}/\partial{J_1}) <0 \quad \text{for}\quad J_{1}>0.
\end{equation}
Numerically solving Eq.\ (\ref{critical}), one obtains that the maximal value
of $h_{cr}$, achieved at $J_1\to 0$, is approximately $h_{cr}\simeq 1.7 J_2$, and
the spin wave velocity in this limit is still of the order of the bandwidth,
$v(h_{cr})\simeq 0.6J_2$, which justifies the applicability of  bosonization
formalism close to $h_{cr}$.  
Within this approach, there is no indication that the chiral phase would be
destabilized by a further increase of the magnetic field, so one may conclude
that it extends from $h_{cr}$ up to the saturation field $h_{s}$.

Recently, influence of strong magnetic fields on a
spin-$\frac{1}{2}$ zigzag chain was studied numerically by means of the
DMRG technique
\cite{Okunishi+99,OkunishiTonegawa03}.  The authors of Refs.\
\cite{Okunishi+99,OkunishiTonegawa03} explain the presence of cusps in the
magnetization curve $m(h)$, observed for large $\alpha=J_{2}/J_{1}$, in terms of
the emergent \emph{two-component} Luttinger liquid phase.  Our findings suggest an
alternative scenario, according to which the phase above the upper cusp
singularity is still described by a \emph{one-component Luttinger liquid},
albeit with a spontaneously broken left-right symmetry. It can be
shown \cite{temovekua} that the cusp itself originates from the Ising
transition at the boundary of the chiral phase, and that the phase below
the lower cusp singularity is also chiral.  Thus, comparing our results with the
DMRG phase diagram \cite{OkunishiTonegawa03},
we can conclude that the TL2 phase in Fig.~1 of Ref.\ \cite{OkunishiTonegawa03}
should be identified as a chirally ordered phase.  For high fields, in
the limit $J_{1} \ll J_{2}$ the stability region of this phase expands with
increasing $J_{1}$, in agreement with our result (\ref{importantresult}). Near
the saturation field, this phase extends up to the classical Lifshits point
$J_2=\frac{1}{4}J_{1}$, which again agrees with our large-$S$ analysis given
below.

\emph{$S=1$ chain.--} One can obtain a bosonized description of a single
(unfrustrated) spin-$1$ chain in magnetic field exceeding the Haldane gap $\Delta$
by accessing the parameters $K$ and $v$ either directly from numerical DMRG
studies \cite{Roncaglia} or, in a more exotic way, from the exact solution of the
integrable $O(3)$ nonlinear $\sigma$-model (NLSM), which itself is believed to provide a
proper effective field-theoretical description \cite{KonikFendley02}.  Then, a
zigzag $S=1$ chain in the regime of strong frustration $J_1\ll J_2$ can be
studied along the same lines as done above for the spin-$\frac{1}{2}$ case, i.e.,
treating the zigzag interaction as a perturbation coupling two LLs.  The LL
parameter of a $S=1$ chain turns out to be \emph{increasing} from the free
fermion value $K=1$ at $h=\Delta$ with the further increase of the field, so
that generally for $h>\Delta$ one has $K>1$ \cite{KonikFendley02,Roncaglia}, 
which resembles a 1D Bose gas \cite{Cazalilla-rev}. This fact leads to a
considerable simplification: coupling two LLs with $K>1$ by a zigzag interchain
coupling yields the same effective field theory (\ref{symantisym}), but
since $K>1$ the only relevant term is the product of chiralities
(\ref{chirprod}). Thus, in contrast to the $S=\frac{1}{2}$
case, a strongly frustrated spin-$1$ chain \emph{immediately} 
enters a chiral phase, described by a
one component Luttinger liquid with spontaneously broken chiral symmetry,
as far as the external field becomes higher than the gap value. Similar
NLSM-based analysis
readily applies to any integer-$S$ zigzag chain 
in the limit of $\alpha \gg 1$ as soon as magnetic field closes the Haldane gap.

\emph{Large-$S$ frustrated chain close to the saturation field.--} 
In the vicinity of the saturation field the emergence
of chirality can be analyzed for an arbitrary spin value $S$. In the coherent
state path integral representation the effective Lagrangian is given by
${\mathcal L}=- S\sum_{n} (1-\cos\phi_{n})\partial_{t}\theta_{n}-\langle
{\mathcal H}\rangle$. Using the ansatz
\begin{equation} 
\label{ansatz} 
(-1)^{n}\sin(\phi_{n}/2)e^{i\theta_{n}} \equiv
\psi_{1,n}e^{i\lambda n} + \psi_{2,n}e^{-i\lambda n},
\end{equation}
one can pass to the continuum limit, treating $\psi_{1,2}$ as smooth fields and
keeping only non-oscillating terms.  For $h$ close to $h_{s}$ the densities of
magnons with momenta around $\pi\pm\lambda$ are small, $|\psi_{1,2}|^{2}\ll 1$,
so one may neglect any terms of a higher than quartic order. After rescaling the
bosonic fields $(2S)^{1/2}\psi_{1,2}\to \psi_{1/2}$, one arrives at the Lagrangian
of the form
\begin{eqnarray} 
\label{2-boson} 
{\mathcal L}&=&\int dx \sum_{\sigma=1,2}
\Big\{ 
i \psi_{\sigma}^{*}\partial_{t}\psi_{\sigma}
-\frac{1}{2m}|\partial_{x}\psi_{\sigma}|^{2} 
+\mu|\psi_{\sigma}|^{2}
\Big\} \nonumber\\
&-&\frac{1}{2}\int dx \{ u (|\psi_{1}|^{2}+|\psi_{2}|^{2})^{2} 
+w|\psi_{1}|^{2}|\psi_{2}|^{2}
\},
\end{eqnarray}
which was recently discussed in the context of 1D binary
Bose-condensate mixtures \cite{CazalillaHo03}.
The Lagrangian parameters are in our case given by
\begin{eqnarray} 
\label{2bpar} 
&&\mu=h_{s}-h,\quad m^{-1}=8J_{2}S\sin^{2}\lambda,\\
&& u=h_{s}/S,\quad w=2\{u-4J_{1}(1+J_{1}^{2}/J_{2}^{2})\sin^{2}\lambda \}.\nonumber
\end{eqnarray}

  In the harmonic fluid approach
\cite{Haldane81} the field operators and densities can be expressed through
scalar bosonic fields $\vartheta$, $\varphi$ as $|\psi_{\sigma}|^{2}=\{\rho_{\sigma}
+\partial_{x}\varphi_{\sigma}/\pi \}\sum_{m}
e^{2im(\pi\rho_{\sigma}x+\varphi_{\sigma})}$ and $\psi_{\sigma}=\{\rho_{\sigma}
+\partial_{x}\varphi_{\sigma}/\pi \}^{1/2}e^{i\vartheta_{\sigma}}\sum_{m}
e^{2im(\pi\rho_{\sigma}x+\varphi_{\sigma})}$, and for $\mu>0$ the Lagrangian
(\ref{2-boson}) describes two LLs of the form
(\ref{SpinChainBosHam}), 
with a density-density interaction.
In contrast to Ref.\ \cite{CazalillaHo03}, in our case the total particle numbers
of the components $n_{1,2}=\int dx |\psi_{1,2}|^{2}$ (which are separately
conserved) are not fixed, but are chosen by the system so as to minimize the energy at
$\mu<0$. It is easy to show that for $w>0$ the system is unstable against any
perturbation making $\rho_{1}\not=\rho_{2}$: indeed, e.g., for
$\rho_{1}>\rho_{2}$ the interaction term leads to renormalization
$\rho_{\sigma}\mapsto
\rho_{\sigma}+\langle\partial_{x}\varphi_{\sigma}\rangle/\pi$ with $\langle
\partial_{x}\varphi_{1}\rangle >\langle\partial_{x}\varphi_{2}\rangle$. 
As a result, the chiral $Z_{2}$
symmetry breaks spontaneously and one of the bands ${\sigma}$ gets depleted.
The effective theory is a 
\emph{single} Luttinger liquid with the parameter $K>1$ depending on the 
dimensionless coupling constant
\begin{equation} 
\label{gamma} 
\gamma=\frac{m u}{\rho_{0}}\simeq \frac{\pi}{2S\sin\lambda} 
\Big(\frac{h_{s}}{4J_{2}S(1-h/h_{s})} \Big)^{1/2},
\end{equation}
where $\rho_{0}=(2\mu
m)^{1/2}/\pi$ is the equilibrium density for small $\mu$ (i.e., in
the vicinity of the saturation field). For $h \to h_{s}$, when $\gamma\gg 1$,  
the LL parameter
tends to $1$ and is given by $K\simeq 1+4/\gamma$, and for $\gamma\ll 1$ (which,
despite the condition $\rho_{0}\ll 1$,
is formally possible for large $S$) one has $K\simeq \pi/\sqrt{\gamma}$
\cite{Cazalilla-rev}. 

The chirality order parameter is directly related to the density difference,
$\kappa\simeq \langle |\psi_{1}|^{2}-|\psi_{2}|^{2}\rangle
\sin\lambda$. Neglecting the depleted
field and using a known expression for the density
correlator \cite{Cazalilla-rev}, 
one obtains the leading asymptotics of the chirality correlation
function:
\begin{equation} 
\label{chir-corr} 
\langle \kappa(x)\kappa(0)\rangle \simeq  
\frac{S^{2}}{\pi^{2}}\Big\{
\frac{h_{s}-h}{J_{2}S}  
-\frac{2K\sin^{2}\lambda}{x^{2}} \Big\}.
\end{equation}
The longitudinal spin correlator $\langle S^{z}(x)S^{z}(0)\rangle$ is also
related to the density and behaves similarly to (\ref{chir-corr}). The leading
part of the transversal spin correlator can be
expressed through $\langle \psi^{\dag}(x)\psi(0)\rangle$ and 
is given by
\begin{equation} 
\label{spin-corr}
\langle S^{+}(0)S^{-}(x)\rangle \simeq 2S\rho_{0} \Big(\frac{K}{\pi \rho_{0}x} 
\Big)^{\frac{1}{2K}} e^{i\lambda x}.
\end{equation}

\emph{Discussion.--} In summary, we have shown that a sufficiently
strong magnetic field applied to a spin-$S$ isotropic $J_{1}$-$J_{2}$ zigzag
chain induces a phase with spontaneously broken $Z_{2}$ symmetry, which is
characterized by the long range vector chirality order and emerges immediately
below the saturation field if the frustration strength $J_{2}/J_{1}$ exceeds the
classical Lifshits point value $\frac{1}{4}$.  This chiral phase is
\emph{gapless} and its low-energy physics is effectively described by a
\emph{one-component} Luttinger liquid. Our results refute the two-component LL
scenario proposed in Refs.\ \cite{Okunishi+99,OkunishiTonegawa03}, and in fact
may necessitate reconsidering the phase diagrams of other frustrated spin
models, particularly of a biquadratic-bilinear spin-$1$ chain in magnetic field
\cite{FathLittlewood}.  One may suppose that the S1 phase in Fig.\ 2
of Ref.\ \cite{FathLittlewood} should have a broken $Z_{2}$ symmetry in the
region beyond the Lifshits point. To clarify this issue we suggest to measure 
directly the chirality correlator $\langle \kappa_{0}^{z}
\kappa_{n}^{z}\rangle$ in the limit $n\to\infty$ above the cusp singularity. 
For spin-$\frac{1}{2}$ chain, such a correlator was calculated only for 
very short distances \cite{Yoshikawa+04} and indicated emergence of at least 
short-range chirality correlations for $h$ directly below the saturation field $h_{s}$.

The chiral phase should be able to survive finite temperature effects since
it involves breaking of the discrete $Z_2$ symmetry. Less trivially, it has also
a chance to survive the three-dimensional interaction without tranforming into a
usual helical long-range order: as noted  by Villain
\cite{Villain78}, at finite temperatures the chirality correlation length is
much larger than the spin correlation length, so with decreasing temperature the
chiral order should set in before the helical spin order does.

Several materials are known which realize zigzag spin-$\frac{1}{2}$ chains (see
Table 1 of Ref.\ \cite{Hase+04}). A promising candidate substance for detecting
the field-induced chirality would be $\rm (N_{2}H_{5})CuCl_{3}$, since its small
exchange constants $J_{1}\simeq 4$~K and $J_{2}\simeq 16$~K make feasible the
task of attaining magnetic fields comparable to $J_{2}$. Experimentally, the
projection of vector chirality $\vec{\kappa}$ on the applied field direction
could be detected by comparing inelastic scattering intensities for oppositely
polarized neutrons, as it was done for the triangular lattice antiferromagnet
$\rm CsMnBr_{3}$ \cite{Maleyev+98}; a similar route can be employed with
polarized light.  We hope that our results will stimulate further experimental
work in this direction.

\emph{Acknowledgments.--}
We thank D. Cabra, A. Honecker, and U. Schollw\"ock for stimulating
discussions. AK is supported by the Heisenberg Fellowship of Deutsche
Forschungsgemeinschaft.


\begin{thebibliography}{10}
\bibitem{comment1}
In fact, (\protect\ref{symantisym}) contains also the 
term $mg_{0}\partial_{x}\phi_{+}$, which however can be absorbed by the
redefinition $\phi_{+} \to
\phi_{+}+(g_0/v)mx$ without affecting our subsequent arguments.
\end{thebibliography}
\end{document}